\documentstyle[11pt,paspconf]{article}
\def\arcdeg{\hbox{$^\circ$}}
\def\ab{$\sim$}
\def\pp{{\it p$-$p}}
\def\lesssim{\mathrel{\hbox{\rlap{\hbox{\lower4pt\hbox{$\sim$}}}\hbox{$<$}}}}

\begin{document}

\title{The High Energy Gamma-Ray Background as a Probe of the Dark
Matter in the Galactic Halo}

\author{R. Chary\altaffilmark{1} \& E. L. Wright}
\affil{Division of Astronomy \& Astrophysics, University of California,
Los Angeles, CA 90095-1562}

\altaffiltext{1}{email: rchary@astro.ucla.edu}

\begin{abstract}
We present constraints on the density of halo dark matter candidates within
the solar circle based on the anisotropy in the high energy gamma-ray
background. The known galactic components of the gamma-ray background, 
in particular the inverse Compton component, have been estimated more
accurately. We find the spectrum of the residual emission, 
after subtracting the
galactic component is inconsistent with emission from some of the
proposed dark matter candidates. We derive upper limits of 10$^{8}$
M$_{\sun}$ for the mass of diffuse gas and 3$\times$10$^{9}$ pc$^{-3}$ for
the number density of primordial black holes contributing to the
gamma-ray background.

\end{abstract}


\section{Introduction}

Various observational results such as those from the Supernova Cosmology
Project, estimates of the primeval deuterium abundances from
primordial nucleosynthesis, microwave background anisotropies,
seem to suggest a partition of the mass and energy
density of the universe into 
$\Omega_{baryon}\approx$0.05, $\Omega_{non-baryonic}\approx$0.3 
and $\Omega_{\Lambda}\approx$0.65. 
Rotation curves of the Galaxy clearly indicates the existence
of a large amount of unseen matter which might be either baryonic or
non-baryonic. For example, the circular velocity {\it v}=220 km/s
at the solar circle R$_{\sun}$=8.5 kpc implies 
an average density ($\rho_{halo}$) 
of 8.4$\times$10$^{-25}$ g/cm$^{3}$. Of this $<$10\% can be
attributed to stars and detectable gas and dust. The nature and
distribution of the remainder of this matter is as yet unknown.
Gamma-rays are a unique probe of the dark matter since they are capable
of tracing both baryonic and non-baryonic dark
matter candidates. Baryonic candidates such as cold, diffuse gas would
produce gamma-rays by interacting with cosmic ray nucleons through the
\pp~process $p_{CR}+p_{gas}\rightarrow p+p+\pi^{0}\rightarrow p+p+2\gamma$. 
Non-baryonic candidates such as WIMPs
which are postulated to be in the mass range 50$-$500 GeV would produce 
line and continuum gamma-rays
through annihilation processes while evaporating primordial black holes
(PBHs)
are thought to produce an E$^{-3}$ photon spectrum with a break below 120 MeV
(Halzen et al. 1991). 
Unfortunately, dark matter candidates are by no means the dominant
source of gamma-rays in the Galaxy. Detecting them
through the gamma-ray background requires accurate estimation of the
other known galactic components of gamma-ray emission such as
nucleon-nucleon (\pp),
inverse Compton (IC) and electron bremsstrahlung (EB) from high energy
cosmic ray electrons scattering off the Coulomb field of the gas.
In addition, there is thought to be an isotropic extragalactic component
from unresolved blazars, \ab50 of which have been detected (Sreekumar et
al. 1998, Mattox et al. 1997). 
The EGRET instrument on the {\it Compton Gamma-Ray
Observatory} has mapped out the entire sky in the energy 
range 30 MeV to 10 GeV with unprecedented sensitivity. 
We have attempted to accurately estimate the galactic component of the
gamma-ray background, especially the IC component, in light
of the recent data available on the interstellar radiation field from
the DIRBE instrument on COBE. We use these new results to 
constrain the nature and the
amount of dark matter in the aforementioned forms within the solar circle. 

\section{Analysis}

The gamma-ray data from the individual EGRET pointings upto April 1997 were
integrated to construct an all sky intensity map 
at 10 energy bands between 30 MeV and 10 GeV. The all sky
map shown in Figure 1 has been generated by co-adding the
maps in the 7 energy bands above 100 MeV and smoothing 
with a 5$\arcdeg$ FWHM gaussian which is approximately the resolution of
EGRET at 100 MeV.
\begin{figure}[!htb]
\plotfiddle{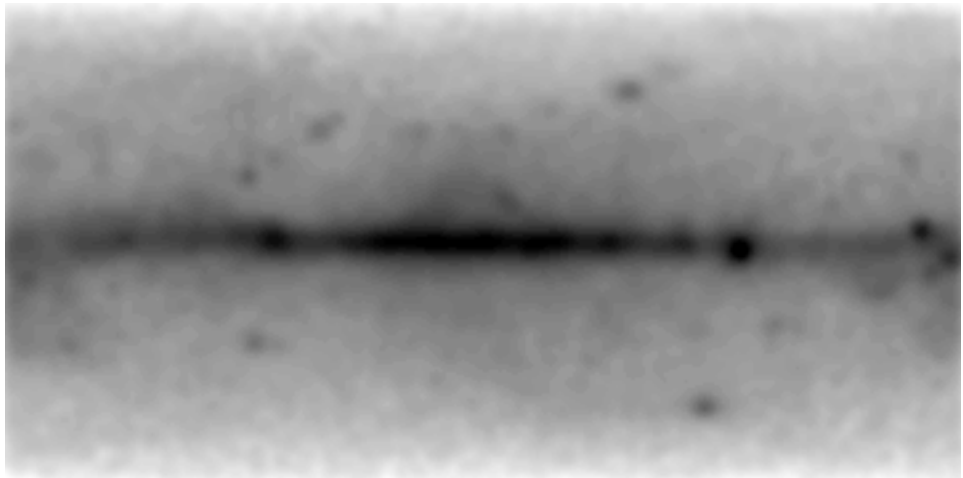}{65truemm}{0}{100}{100}{-305}{-300}
\caption{The E$>$100 MeV gamma-ray all sky map reconstructed from 
individual EGRET pointings.} 
\end{figure}
Immediately noticeable is a strong component of gamma-ray emission aligned 
with the galactic plane which is though to arise from the \pp~process
described above. Since the scale height of the neutral hydrogen gas 
in the Galaxy
is of order 100 pc, we can eliminate most of the galactic component by
restricting our analysis to high latitudes i.e. $|b|>30\arcdeg$.  
Arendt et al. (1998) performed an analysis comparing the DIRBE 100$\micron$ 
intensity to the column density of gas 
in the Galaxy and determined a good correlation between the two. 
Therefore, to subtract off the high latitude \pp~component, 
we performed a linear correlation of the form I$_{\gamma}$=B~I$_{100\mu}$
+ C, between the DIRBE
100$\micron$ map and the EGRET 100$-$150 MeV intensity map. The constant
B was determined to be 1.0$\times$10$^{-8}$ photons cm$^{-2}$
s$^{-1}$ sr$^{-1}$ MeV$^{-1}$ per MJy/sr of the 100$\micron$ map. 
In comparison, Bertsch et al. (1993) 
adopted an analytical value
of 5$\times$10$^{-25}$ photons s$^{-1}$ GeV$^{-1}$ nucleon$^{-1}$ 
for the gamma-ray intensity per H atom at 125 MeV. Assuming the 
value of 18.6 nW m$^{-2}$ sr$^{-1}$/10$^{20}$ cm$^{-2}$ determined by 
Arendt et al. (1998) for the ratio of the 
100$\micron$ intensity to the gas column density, 
we derive a value of 6$\times$10$^{-9}$ gamma-ray
photons cm$^{-2}$ s$^{-1}$ sr$^{-1}$ MeV$^{-1}$ per MJy/sr of 
the DIRBE 100 $\micron$ map which is consistent with our fit
parameters.
After performing the correlation between one of the EGRET energy bands,
the contribution from the gas in the Galaxy is scaled to the 
other EGRET energy bands using the
well determined \pp~and EB source function (Bertsch et al.
1993).

The IC contribution to the gamma-ray background has been
rather uncertain. The process involves scattering of high energy cosmic
ray electrons in the energy range 0.1 GeV to 1 TeV off starlight, 
dust reprocessed infrared emission and the cosmic microwave background.
The only detailed modelling of this component 
has been by Bloemen (1985) who assumed an isotropic distribution for
both the cosmic ray electrons and the interstellar radiation field
(ISRF). An isotropic assumption for the ISRF is 
clearly inaccurate outside the galactic plane because the scale height
of the stars and dust is of order 100 pc. We have reconstructed a model
for the ISRF in the wavelength range 0.1$\micron$ to 1000$\micron$
based on work done by Mathis et al. (1981). The model was 
fit to the
DIRBE all sky maps from 1.2$\micron$ to 240$\micron$. For wavelengths
shorter than 1.2$\micron$, we adopt the average intensity values from
the TD1 and the Pioneer 10 missions (See Chary \& Wright 1998 for
details about the ISRF model and the characteristics of the IC
emission). The IC component is not very sensitive to the values
at these shorter wavelengths since $\nu$J$_{\nu}$ is lower by a factor of
5 compared to the peak at 1.2$\micron$. 
The ISRF model is then used to calculate
the intensity of radiation at each point within the Galaxy
as a function of solid angle. This is convolved with a cosmic ray
electron distribution which is assumed to be isotropic and the
Klein-Nishina scattering cross-section to obtain the gamma-ray source
function. 
\begin{equation}
\int_{30 MeV}^{10 GeV}j_{\nu,\gamma,r,z} d\nu =
\int_{0.1\micron}^{1000\micron}
\int_0^{4\pi} \int_0^{4\pi} \int_{0.1 GeV}^{1 TeV}
\frac{I_{\nu,r,z}(\Omega)}{c}
\frac{dn}{dE} \frac{d\sigma_{KS}}{d\Omega} dE d\Omega d\Omega d\nu
\end{equation}
Here $j_{\nu,\gamma,r,z}$ is the IC gamma-ray source
function
as a function of galactocentric radius ({\it r})
and height from the plane ({\it z}),
I$_{\nu,r,z}(\Omega)$ is the intensity of the ISRF as a function of
solid angle, $\frac{dn}{dE}$ is the cosmic ray electron intensity and
$\frac{d\sigma_{KS}}{d\Omega}$ is the Klein-Nishina scattering cross
section. For this paper, we adopt a cosmic ray electron distribution
which has a spectral shape similar to that at the solar neighbourhood 
but varies spatially as:
\begin{equation}
\frac{dn}{dE}(r,z) = (\frac{dn}{dE})_{\sun} e^{-z/750} (1+(R_{\sun}-r)/5000)
\end{equation}
Chary \& Wright (1998) however, report on the IC
contribution from different cosmic ray electron distributions.
The source function is integrated along all lines of sight 
within the Galaxy to
obtain an IC intensity map for high latitudes 
in the ten EGRET energy bands. Then,
the final EGRET intensity maps have any bright point sources masked,
the \pp, EB and IC components subtracted and the
residue analyzed for excess emission suggestive of a gamma-ray halo.

\section{Results and Dark Matter Constraints}

Many authors (e.g. Dixon et al. 1998) have suggested the existence of 
a halo of gamma-ray emission surrounding the Galactic center 
with intensity \ab10$^{-6}-10^{-7}$ photons cm$^{-2}$ s$^{-1}$ sr$^{-1}$
above 1 GeV.
This could be attributed either to:
\begin{enumerate}
\item a halo dark matter component or 
\item the inverse-Compton component or 
\item a population of unresolved
sources such as pulsars, in the direction of the galactic center. 
\end{enumerate}
We find that the magnitude of this halo intensity is similar to the
difference in the IC intensity between l\ab0$\arcdeg$ and
l\ab180$\arcdeg$ which is of order 10$^{-7}$ photons cm$^{-2}$ s$^{-1}$
sr$^{-1}$ above 1 GeV.
The anisotropy in the gamma-ray background before and after subtraction of 
the IC component is shown in Figure 2. 

\begin{figure}[!htb]
\plotfiddle{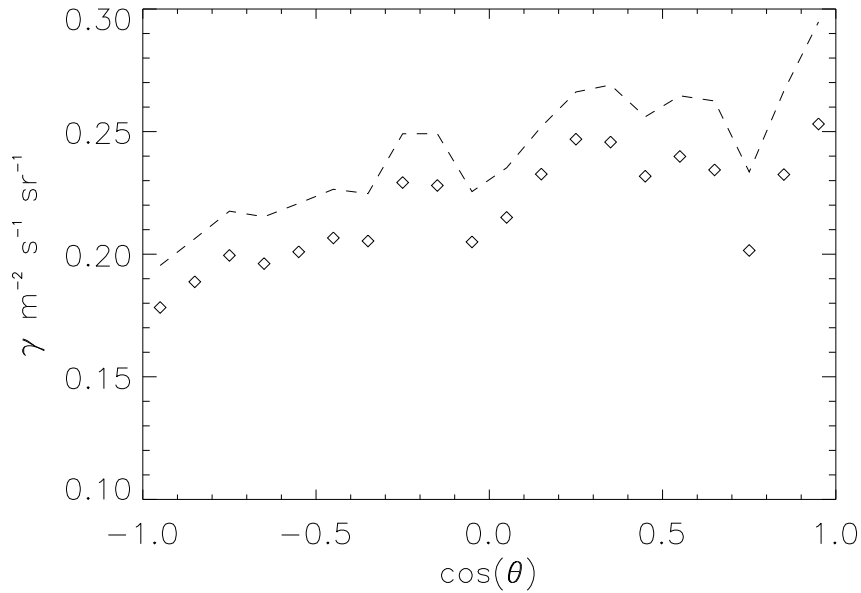}{75truemm}{0}{70}{70}{-240}{-270}
\caption{The latitude averaged ($|b|=$30\arcdeg$-$60$\arcdeg$)
profile of the gamma-ray intensity
in the energy range 70 MeV$-$4 GeV 
as a function of $\theta$. $\theta$ is the angle between l=0$\arcdeg$
and the line of sight. The line indicates the profile with only the \pp~
and EB components subtracted while the symbols represent the profile
after the IC component has been subtracted.}
\end{figure}

Clearly, some of the
anisotropy in the gamma-ray maps after the \pp~and EB components are
subtracted, can be attributed to the IC
component. However, there still appears to be a enhancement of order
15\% in the gamma-ray intensity towards l\ab0$\arcdeg$ after 
subtraction of the IC component. 
The differential photon spectrum of this residual emission 
above 100 MeV is E$^{-1.8}$
with an intensity of 2$\times$10$^{-6}$ photons cm$^{-2}$
s$^{-1}$ sr$^{-1}$, an order of magnitude less than the isotropic
background.
This spectrum is {\it inconsistent} 
with emission from the \pp~process, PBHs
or annihilating WIMPs (Gondolo 1998). 

A fit to the residual spectrum provides
an upper limit of 3$\times$10$^{9}$ pc$^{-3}$ (H$_{0}$=65 km s$^{-1}$
Mpc$^{-1}$) to the number
density of PBHs in the Galactic halo within the solar circle. 
However, since the PBHs that dominate 
the diffuse gamma-ray emission have mass of order 10$^{14}$~g, 
it is not
possible to constrain $\rho_{PBH,halo}$ without knowing the 
mass spectrum
of the PBHs at the time of formation. For example, if PBHs formed in the
radiation dominated era resulting in a
$\frac{dn}{dM}\propto M^{-2.5}$ spectrum,
the derived average halo mass density from PBHs would be
$\rho_{PBH,halo}<2\times$10$^{-7}$ $\rho_{halo}$ within the solar
circle. 

Alternatively, if the excess gamma-ray emission is attributed to 
the \pp~ component from diffuse high latitude gas, the derived maximum
column density is \ab5$\times$10$^{19}$ cm$^{-2}$. This is assuming a
gamma-ray emissivity per H atom similar to that used to subtract the 
Galactic \pp~component earlier. The upper limit to the total mass of
diffuse halo gas within the solar circle is then 
$\lesssim 10^{8}$~M$_{\sun}$.

Based on the spectrum of the residual emission, we conclude that the 
enhancement
is more likely to be due to an underestimation of the inverse 
Compton component which in
our simulations has a photon index of $-$2.0. This
implies
a cosmic ray electron density which is steeper than the
adopted distribution, in the direction of the galactic center and at
high latitudes.
Another possible interpretation is a population of unresolved point
sources such as pulsars. We estimate that the residual emission can be 
produced by about 100 pulsars with Crab like gamma-ray intensities.
Better spatial resolution data which will be provided in the future by the GLAST
gamma-ray mission is required to test this hypothesis.



\begin{references}
\reference Arendt, R. G., et al., 1998, \apj, in press
\reference Bertsch, D. L., et al., 1993, \apj, 416, 587
\reference Bloemen, J. B. G. M., 1985, \aap, 145, 391
\reference Chary, R., \& Wright, E. L., 1998, in preparation for ApJ
\reference Dixon, D., et al., 1998, New Astronomy, 3, 7, 539
\reference Gondolo, P., 1998, astro-ph/9807347
\reference Mathis, J. S., Mezger, P. G., \& Panagia, N., \aap, 128, 212
\reference Mattox, J. R., et al., 1997, \apj, 481, 95
\reference Sreekumar, P., et al. 1998, \apj, 494, 523
\reference Halzen, F., et al., 1991, Nature, 353, 807

\end{references}
\end{document}